# Confidently Wrong: Severity-Aware Calibration of Prompt-Injection Detectors under Attack Shift

Md Anas Biswas
*School of Computing, University of Portsmouth, Portsmouth, United Kingdom*
up2082724@myport.ac.uk
ORCID: 0009-0009-0113-5816

## Abstract

Prompt-injection detectors are deployed as guards: a model scores an input and a downstream system trusts or blocks it on that score. I study the confidence of these scores, not only their accuracy, when the attack distribution shifts away from the clean benchmark on which the operating point was chosen. I evaluate three released detectors, ProtectAI-v2 and two Prompt-Guard-2 checkpoints, at a single source-calibrated threshold that I freeze and transport across five shifts. I report a severity metric S, how confident a detector is on the attacks it misses, alongside the false-negative rate and discrimination. Across every shift and every detector, severity on the missed attacks stays between 0.99 and 1.00 while the false-negative rate ranges from 0.01 to 0.97: when these detectors miss, they miss with near-certainty. All three confidently pass indirect behavior-hijack injection, a blind spot unanimous across two vendors and a fourfold size range. Standard pooled calibration error does not register this; one detector it rates well-calibrated, at 0.06, is miscalibrated at 0.91 on the attacks alone. Run against live models, the missed injections leak the majority of working exploits, passing them at the rate they catch others. A controlled experiment traces the cause to content-keying rather than injection structure, an instruction-tuned model used as a judge shows the same hijack blind spot, and a black-box rewriter exploits the content-keying to manufacture working confident misses, most effectively on the most dangerous attack category. Code and data are public.

**Keywords:** *prompt injection; indirect prompt injection; guard models; LLM security; calibration; confident false negatives; distribution shift; adversarial robustness*

## 1. Introduction

Applications built on large language models routinely place untrusted text in front of the model: retrieved documents, tool outputs, emails, and web pages. Prompt injection, in which an instruction hidden in that untrusted text redirects the model away from its intended task, is the dominant practical attack against such applications. A common defense is to place a detector, often called a guard model, in front of the application. The detector scores an input and the application uses that score to decide whether to trust, flag, or block the input.

The score is treated as a confidence. A low attack-probability is read as evidence that the input is safe, and the application proceeds. This reading rests on an assumption that is rarely measured directly: that the detector is calibrated, so that a low score corresponds to a genuinely low chance of attack. Public benchmarks for these detectors report accuracy on clean, in-distribution splits, typically F1 and area under the ROC curve. Deployment does not stay in distribution. New attack styles appear, attacks move from standalone prompts into documents and tool outputs, and the operating point chosen on a clean benchmark is carried unchanged into a shifted world.

The risk is asymmetric. A high-confidence false negative, in which the detector assigns a real attack a low attack-probability with near-certainty, is worse than a low-confidence one, because the application trusts it more. A pooled calibration error, computed over attacks and benigns together, averages this danger away: it is dominated by the benign majority and does not isolate the confidence of the misses that matter. What a security operator needs to know is narrower, namely how confident the detector is on the attacks it lets through, after the distribution has shifted under a frozen operating point. Figure 1 shows this pipeline and the confident-miss path it creates.

A guard score is trusted as a safety confidence. The dangerous event is not a miss, but a confident miss.
Untrusted input (document, email, tool output)
Guard detector scores p = P(attack)
Frozen threshold t (set on clean benchmark)
p ≥ t : block
p < t : pass to downstream LLM
LLM acts on the input
Confident false negative: a real attack scored p ≈ 0 (p < t with near-certainty)
severity S = mean (1 - p) on these misses → near 1

*Figure 1. The deployment pipeline. A guard detector scores an untrusted input, and a frozen threshold set on a clean benchmark decides whether the input is blocked or passed to the downstream model. The failure this paper measures is the highlighted path: a real attack scored below the threshold with near-certainty, a confident false negative, which the downstream system trusts as a safe verdict. The severity metric S measures the confidence on exactly these misses.*

This paper makes six contributions.

1. A security-asymmetric severity metric, S, that scores the confidence of false negatives specifically. S is attack-conditional and is invariant to the benign base rate by construction, so it measures an intrinsic property of the detector rather than an artifact of class balance.
2. A frozen-threshold transport panel that fixes one operating point per detector on a clean direct benchmark and measures three released detectors across five shifts, with bootstrap confidence intervals on every cell.
3. A proposition showing that the standard pooled calibration error cannot bound this risk, since confident false negatives are diluted into benign-dominated bins as the base rate falls, together with evidence that a detector rated well-calibrated by the pooled metric is, on the attacks alone, severely miscalibrated on the same scores.
4. Downstream validation: the confidently-missed injections are run against live target models, and the detectors leak the majority of the attacks that actually succeed, passing them at the same rate as the attacks they catch.
5. A controlled mechanism experiment showing that the detectors key on harmful surface content rather than on injection structure, so embedding an injection in benign content suppresses or fails to raise the score, together with a generative-judge detector that shares the maximal severity and the hijack blind spot yet adds an instruction-following attack surface the encoders lack.
6. A black-box severity-maximizing adversary that exploits the content-keying to rewrite genuine injections into innocuous prose, driving the detectors' miss rate close to one and, after a downstream functionality check, manufacturing working confident misses most effectively on the most dangerous attack category.

I state the scope of the claim plainly. I do not claim to be the first to study calibration of guard models, nor the first to observe informally that a low pooled calibration error fails to bound a security risk. The contributions I claim are the proposition making that failure precise, the security-asymmetric severity lens applied across a frozen-threshold transport panel, the downstream-validated exploitability of the confident misses, the structural mechanism behind them, and the adversary that weaponizes it.

# 2. Related work

***Detectors and benchmarks.***

Released prompt-injection detectors include ProtectAI's DeBERTa-based classifier [14] and Meta's Prompt-Guard family [15], which I evaluate here. Standard datasets cover distinct slices of the problem: deepset [16] supplies direct standalone-prompt injections and benigns, BIPIA [7] supplies indirect injection, the threat introduced by Greshake et al. [19], where the attack is an instruction embedded in a host document, a jailbreak-classification set [17] supplies standalone roleplay and jailbreak prompts as a near in-distribution stress test, and NotInject supplies benign prompts laced with injection trigger words to probe over-defense, in the line of work introduced with InjecGuard [6]. These datasets each anchor one shift in my panel.

***Evaluation methodology under shift.***

Recent methodology work argues that standard splits overstate detector quality. The Gate AI evaluation [3] fixes a global operating point at a low false-positive rate, deduplicates across datasets, and uses leave-one-dataset-out testing to measure threshold transferability. Concurrent work on benchmark reliability [2] reports that standard splits inflate area under the curve by several points relative to leave-one-dataset-out, that a large fraction of features are dataset shortcuts, and that several production detectors fail on indirect attacks. I adopt the global frozen operating point and the transport framing from this line, and I differ from it by scoring the confidence of false negatives rather than accuracy or F1.

***Calibration of guard models.***

The closest calibration work [1] studies the expected calibration error of large-language-model guard models and reports degradation under jailbreak conditions. Separate work [4] argues that a low expected calibration error does not bound downstream risk, because the pooled error is insensitive to where in the score range the miscalibration sits. My severity metric is consistent with both observations and adds a security-asymmetric, attack-conditional view that the pooled error does not provide.

***Temporal drift, robustness, and selective prediction.***

Other studies track how false-negative rates of deployed filters drift upward over months as attacks evolve [5]. Adaptive-attack-aware detection has been studied through known-answer and minimax training [8]; my adversary is complementary and deliberately weaker, a black-box paraphraser with no access to detector internals, used to show that the content-keying alone is exploitable. Work on selective prediction and risk-coverage analysis [20] provides the abstention framing I draw on in the discussion. Calibration background, including temperature scaling [10] and calibration under dataset shift [11], and the lineage of calibration-error estimators [12, 13], informs the metric design. My position relative to this work is to combine the security-asymmetric confidence-of-false-negatives lens with the frozen-threshold transport protocol, and to report the result as a panel with bootstrap intervals across detectors, shifts, and model size.

# 3. Method

## 3.1 Problem setup and detection unit

I study released detectors operated as confidence sources. Each detector takes an input and returns a continuous score p equal to its estimated probability of attack, which a downstream system may threshold to block, flag, or trust the input. The failure I focus on is the high-confidence false negative: the detector assigns a real attack a low attack-probability, the downstream system reads that low score as strong evidence of safety, and the attack passes with the detector's endorsement.

I fix the detection unit explicitly, because it determined the validity of earlier measurements, discussed in Section 5. For direct and jailbreak prompts the unit is a standalone prompt. For indirect injection the unit is an instruction embedded inside a host document, because that is the form the attack takes in deployment. Scoring a bare injected instruction in isolation, with no host document, measures something other than indirect-injection detection.

## 3.2 Datasets and shifts

Table 1 lists the datasets, the shift each anchors, the detection unit, the counts after assembly and deduplication, and the role each plays. A shift is a target distribution against which the frozen detector is measured.

| Dataset | Shift | Unit | Attacks | Benigns | License | Role in study |
|---|---|---|---|---|---|---|
| deepset/prompt-injections | direct | standalone prompt | 263 | 399 | Apache-2.0 | source for threshold; in-distribution baseline |
| microsoft/BIPIA | indirect-harmful | instruction in document | 60 | 778 | MIT | scams and misinformation injected in email/table documents |
| microsoft/BIPIA | indirect-hijack | instruction in document | 150 | 778 | MIT | encoding, cipher, reverse, emoji, translation behavior overrides |
| jackhhao/jailbreak-classification | jailbreak | standalone prompt | 396 | 398 | Apache-2.0 | near in-distribution stress test |
| leolee99/NotInject | over-defense | standalone benign | 0 | 339 | MIT | benign trigger-word prompts; over-defense rate |

*Table 1. Datasets and shifts. Benign counts for the two indirect rows are the same BIPIA host-document pool. Licenses are taken from each dataset card: deepset and jackhhao under Apache-2.0, NotInject and BIPIA under MIT. All datasets are used within their licensed terms for research evaluation.*

The indirect shifts are assembled from BIPIA host documents, drawn from its email and table tasks, with a single injected instruction appended at the end of the document. I separate BIPIA's text-attack categories into two tiers. The harmful tier, comprising Scams and Fraud and Misinformation and Propaganda, inserts content whose presence is itself damaging. The hijack tier, comprising Base Encoding, Substitution Ciphers, Reverse Text, Emoji Substitution, and Language Translation, overrides the model's intended output behavior without inserting harmful content, which is the structural definition of a successful injection. I drop the remaining BIPIA categories, which are benign task instructions such as recommending a book or summarizing a report, that a guard model is arguably right to score low; this decision is justified empirically in Section 5. I deduplicate within and across sources before scoring; the cross-source near-duplicate overlap is zero (Appendix Table A2), so the transport gaps reflect genuinely distinct distributions rather than shared examples.

## 3.3 Detectors

I evaluate three released detectors, used as published, with no retraining:

- ProtectAI-v2, a DeBERTa-v3-base sequence classifier (protectai/deberta-v3-base-prompt-injection-v2), with labels SAFE and INJECTION.
- Prompt-Guard-2 at 86M parameters (meta-llama/Llama-Prompt-Guard-2-86M).
- Prompt-Guard-2 at 22M parameters (meta-llama/Llama-Prompt-Guard-2-22M).

The two Prompt-Guard checkpoints come from one vendor and one family and differ by roughly four times in parameter count, which gives a controlled capacity axis. ProtectAI-v2 is a second vendor and a different architecture. For every detector I read p as the softmax probability of the attack class, identified from the model's label map, and I truncate inputs to 512 tokens.

## 3.4 Frozen-threshold transport protocol

I calibrate one operating point per detector and freeze it. For each detector I set the threshold t at the 0.99 quantile of the detector's scores on the direct benign distribution, which targets a one percent false-positive rate in-distribution. I then apply that same t to every shift without recalibration. This isolates the transport question: given an operating point a practitioner would reasonably choose on a clean direct benchmark, how does the detector behave when the attack distribution moves under it.

## 3.5 Metric family

All rates are attack-conditional. Let the number of attacks be the attack count, let p be the attack-probability, let t be the frozen threshold, and let a miss be the event that the input is an attack and p is below t.

- False-negative rate (FNR): the fraction of attacks that are missed. The standard accuracy view.
- Severity (S): the mean value of one minus p across the missed attacks. The headline metric. S close to one means the misses are near-certain rather than borderline. Reported as not estimable when a cell has fewer than ten misses.
- Attack-conditional calibration error (ECE_atk): the expected calibration error of p restricted to true attacks, taken against the constant true label of one that every attack carries, so it equals the binned mean distance of p from one over the attacks. It is weight-free and base-rate-invariant. Severity, the attack-class Brier score, and ECE_atk are three aggregations of one underlying quantity, the benign-confidence the detector places on attacks: a conditional mean over the misses, a squared mean, and a binned reliability over all attacks; they agree by construction and are reported together as a coherent family rather than as independent corroboration. I report ECE_atk against the standard pooled calibration error (ECE_pooled), which is kept only for contrast.

- Attack-class Brier score: a proper scoring rule restricted to attacks, the mean squared distance of p from one over the true attacks.
- Benign false-positive rate and AUROC: the over-defense rate and threshold-free discrimination, reported wherever the relevant classes are present.

The headline severity metric and its two attack-conditional companions are defined as follows, with N-plus the number of attacks, M the set of missed attacks, and B_b the attacks whose score falls in bin b.

$$S = \frac{1}{|M|} \sum_{i \in M} (1 - p_i), \quad M = \{\, i : \ y_i = 1, \ p_i < t \,\}$$

$$\mathrm{ECE}_{\mathrm{atk}} = \sum_{b} \frac{|B_b|}{N^+} \left| 1 - \frac{1}{|B_b|} \sum_{i \in B_b} p_i \right|$$

$$\mathrm{Brier}_{\mathrm{atk}} = \frac{1}{N^+} \sum_{i:\, y_i = 1} (1 - p_i)^2$$

S, the false-negative rate, and the over-defense rate are computed from frozen scores at a fixed t, so they depend only on the detector's behavior on attacks, or on benigns, and not on the benign-to-attack prevalence. They are therefore invariant to the base rate by construction. Figure 2 verifies this invariance on synthetic prevalence sweeps: as the benign prevalence varies across several orders of magnitude, S and the attack-conditional rates stay flat while a pooled calibration error moves. Deployed risk, which a practitioner would obtain by weighting the false-negative rate and severity by the operational base rate, is prevalence-dependent on purpose and is not the subject of the panel.

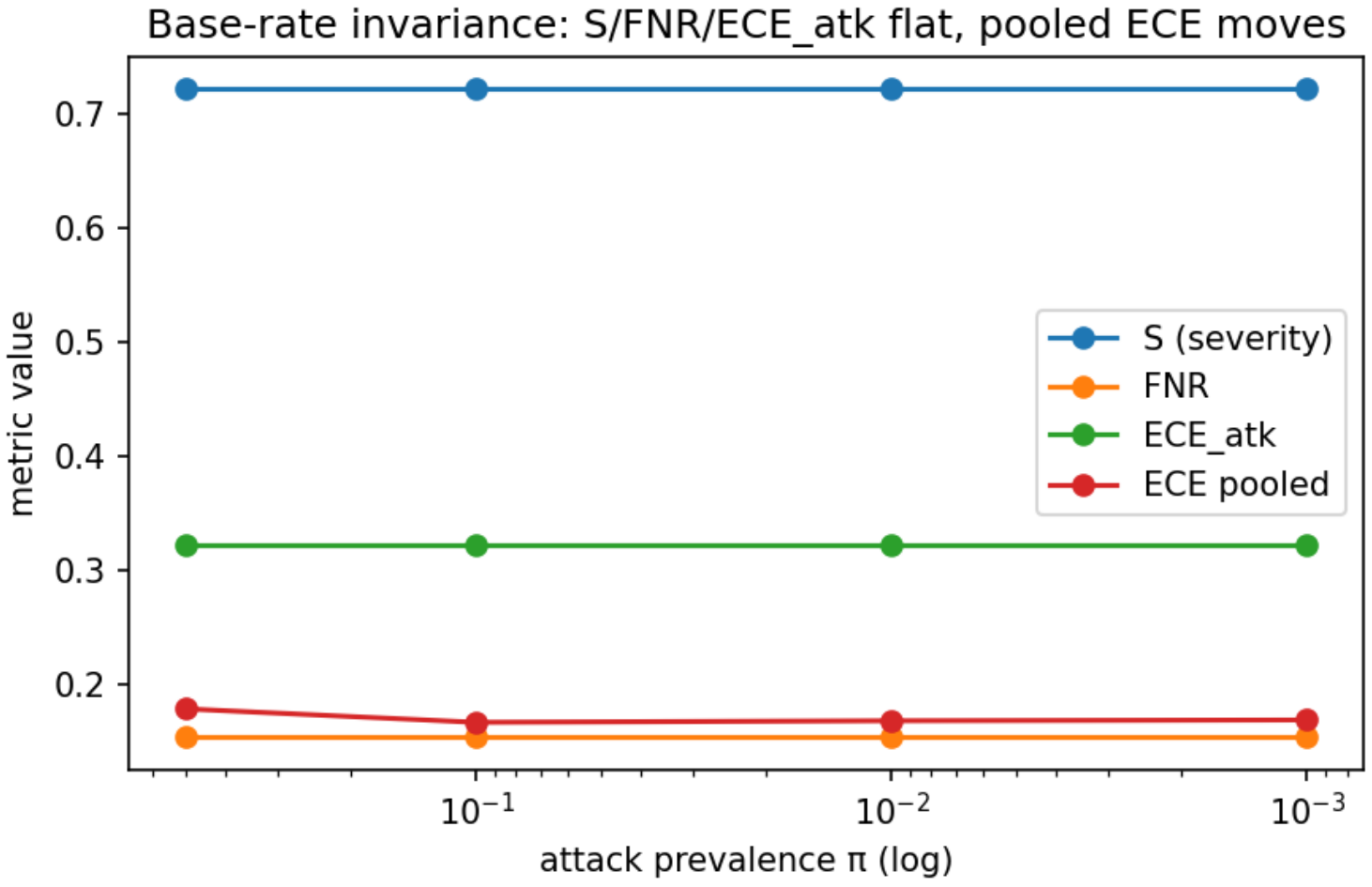

*Figure 2. Base-rate sweep on a synthetic example, illustrating the invariance property rather than a measured detector. With scores and the threshold frozen, the attack-conditional metrics including severity stay flat as benign prevalence varies, while a pooled calibration error moves. The vertical positions are properties of this synthetic example, not of any detector in the panel; the point is the flatness of the attack-conditional curves*

*against the drift of the pooled one. Severity reports an intrinsic property of the scores, not an artifact of test-set balance.*

### *3.5.1 Why pooled calibration error does not bound this risk*

Severity is motivated by a property of the pooled expected calibration error that I state precisely. Write the pooled error as a population-weighted average over score bins, the sum over bins b of (n_b / N) times the absolute gap between conf_b and acc_b, where n_b is the number of evaluation samples in bin b, N is the total, conf_b is the mean predicted attack-probability in the bin, and acc_b is the empirical fraction of true attacks in the bin. Severity is the mean of one minus p over the missed attacks, conditioned on attacks alone.

$$\mathrm{ECE}_{\mathrm{pooled}} = \sum_{b} \frac{n_b}{N} \left|\mathrm{conf}_b - \mathrm{acc}_b\right|$$

**Proposition.** Consider a set of m confidently missed attacks, each a true attack with predicted attack-probability at most a small value e. Then the severity over these misses is at least one minus e, while their total contribution to the pooled calibration error is at most m / N. Consequently, as the attack base rate falls, that is, as N grows with m fixed, the contribution of these misses to the pooled error tends to zero while their severity stays at least one minus e. A detector can therefore have a pooled calibration error arbitrarily close to zero while remaining maximally severe on the attacks it misses.

$$S \geq 1 - e \qquad \text{and} \qquad \sum_{b} \frac{n_b}{N} \left|\Delta \mathrm{acc}_b\right| \leq \frac{m}{N}$$

***Proof.*** Each missed attack has p at most e, so it contributes one minus p, at least one minus e, to the severity average, which gives the first part. For the second part, these attacks fall in the low-score bins, whose mean confidence conf_b is at most e. Adding the m attacks across those bins raises each bin's empirical attack fraction acc_b by the count of attacks placed in it divided by n_b; weighting the induced gap by n_b / N and summing across the affected bins, the total contribution is bounded by the count of such attacks divided by N, that is m / N, since each contributes weight 1 / N and a gap of at most one. The benigns, which are correctly scored low, share these same low bins, so as they grow in number N grows, m / N falls, and the confident misses are absorbed into benign-dominated bins. Severity, conditioned on the attacks, does not dilute, because removing the benigns from the average leaves it unchanged.

Two consequences follow. First, the pooled error is least sensitive to confident false negatives in exactly the regime that matters operationally, where attacks are rare and benigns dominate, because that is the regime in which m / N is smallest. Second, the pooled error is symmetric in the sign of the miscalibration: it scores a confident false negative and an equally confident false alarm identically, so it cannot express the asymmetry that makes a trusted missed attack the costly event. Severity is the attack-conditional, sign-aware companion that the pooled error provably cannot replace. Section 4.6 is the empirical instance of this proposition: Prompt-Guard-2 on harmful indirect injection has a pooled error of 0.062, which passes a standard calibration audit, and an attack-conditional error of 0.907 on the same scores.

## 3.6 Statistics

For every cell I compute bootstrap 95 percent confidence intervals on the false-negative rate and on severity from one thousand resamples of the attack rows, taking the 2.5 and 97.5 percentiles.

Where severity is not estimable, with fewer than ten misses, I suppress both its point estimate and its interval, so that no severity figure rests on too few events.

### 3.7 Downstream attack-success validation

A miss is a disagreement with a label until I show the missed attack actually works against a model. I therefore run each genuine indirect injection through live target models and measure whether the model complied with the injected instruction. The targets are two open instruct models, Qwen2.5-7B-Instruct and Qwen2.5-3B-Instruct [9], run in four-bit precision. Each injected document is presented to the target as external content to summarize, with no injection defense in the system prompt, because in my threat model the detector is the defense. I decide attack success programmatically where a reliable surface signal exists: a decodable base64 blob for Base Encoding, an emoji for Emoji Substitution, a non-English output for Language Translation, and the injected scam keyword or link for Scams and Fraud. For the remaining categories I use a separate language model as a judge with a strict yes-or-no rubric. To avoid counting behavior the model produces spontaneously, I run the same target on the benign host documents without injection and subtract that baseline rate. I then join attack success with each detector's confident misses.

### 3.8 Structural-versus-content experiment

To isolate why the detectors miss indirect injections, I run a controlled two-by-two. The first factor is payload, varied between harmful-worded instructions and benign-worded behavior-hijack instructions. The second factor is structure, varied between a standalone prompt and the same instruction embedded in a fixed set of host documents. Holding the host documents and the instruction set constant, a content-keyed detector responds to the payload factor and ignores the structure factor, while a structure-keyed detector responds to embedding regardless of payload. The decisive cell is a benign-worded instruction embedded in a document, which is a genuine injection with a benign payload.

## 4. Results

Table 2 reports the full panel of three detectors crossed with five shifts, with bootstrap confidence intervals. Figure 3 renders severity and the false-negative rate as heatmaps.

| Detector | Shift | t | FNR [95% CI] | S [95% CI] | misses | ben. FPR | AUROC |
|---|---|---|---|---|---|---|---|
| ProtectAI-v2 | direct | 0.028 | 0.548 [0.483, 0.605] | 0.999 [0.998, 0.999] | 144 | 0.010 | 0.882 |
| ProtectAI-v2 | indirect-harmful | 0.028 | 0.650 [0.533, 0.767] | 0.995 [0.993, 0.997] | 39 | 0.397 | 0.444 |
| ProtectAI-v2 | indirect-hijack | 0.028 | 0.693 [0.620, 0.767] | 0.993 [0.992, 0.995] | 104 | 0.397 | 0.424 |
| ProtectAI-v2 | jailbreak | 0.028 | 0.136 [0.104, 0.169] | 0.995 [0.993, 0.997] | 54 | 0.013 | 0.986 |
| ProtectAI-v2 | over-defense | 0.028 | n/a | n/a | 0 | 0.460 | n/a |
| Prompt-Guard-2 (86M) | direct | 0.004 | 0.532 [0.468, 0.593] | 0.999 [0.999, 0.999] | 140 | 0.010 | 0.942 |
| Prompt-Guard-2 (86M) | indirect-harmful | 0.004 | 0.217 [0.117, 0.333] | 0.998 [0.997, 0.998] | 13 | 0.145 | 0.894 |
| Prompt-Guard-2 (86M) | indirect-hijack | 0.004 | 0.693 [0.620, 0.767] | 0.998 [0.998, 0.998] | 104 | 0.145 | 0.625 |
| Prompt-Guard-2 (86M) | jailbreak | 0.004 | 0.010 [0.003, 0.020] | n/a | 4 | 0.166 | 0.993 |
| Prompt-Guard-2 (86M) | over-defense | 0.004 | n/a | n/a | 0 | 0.192 | n/a |
| Prompt-Guard-2 (22M) | direct | 0.021 | 0.837 [0.791, 0.878] | 0.996 [0.995, 0.996] | 220 | 0.010 | 0.777 |

| Detector | Shift | t | FNR [95% CI] | S [95% CI] | misses | ben. FPR | AURO C |
|---|---|---|---|---|---|---|---|
| Prompt-Guard-2 (22M) | indirect-harmful | 0.021 | 0.900 [0.817, 0.967] | 0.996 [0.995, 0.996] | 54 | 0.027 | 0.694 |
| Prompt-Guard-2 (22M) | indirect-hijack | 0.021 | 0.967 [0.933, 0.993] | 0.996 [0.996, 0.997] | 145 | 0.027 | 0.585 |
| Prompt-Guard-2 (22M) | jailbreak | 0.021 | 0.083 [0.058, 0.114] | 0.990 [0.988, 0.992] | 33 | 0.196 | 0.955 |
| Prompt-Guard-2 (22M) | over-defense | 0.021 | n/a | n/a | 0 | 0.130 | n/a |

*Table 2. Transport panel. The column t is the frozen per-detector operating point. Brackets are bootstrap 95 percent intervals. Severity is suppressed where misses are below ten.*

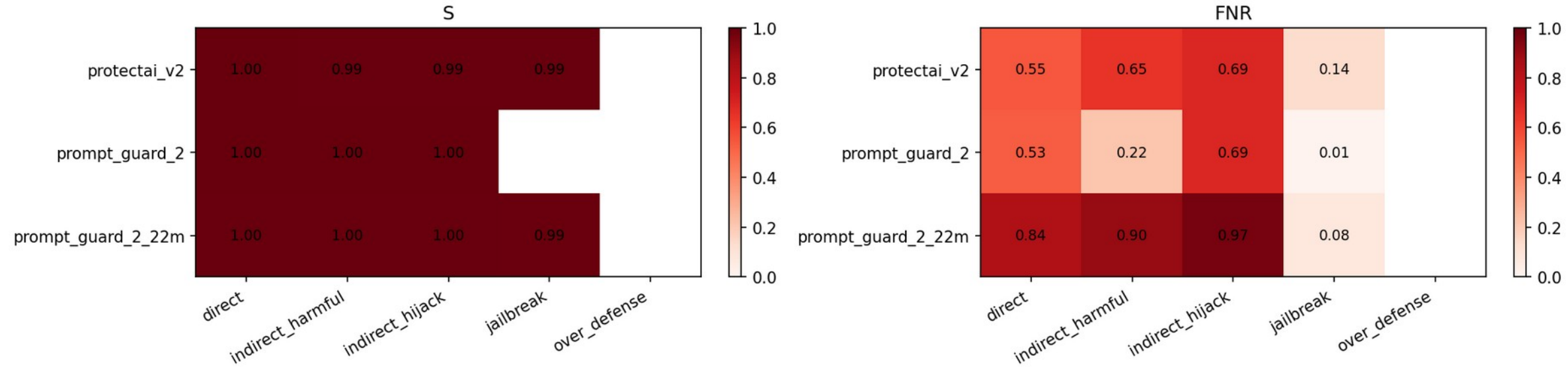

*Figure 3. Severity (left) and false-negative rate (right) across detectors and shifts. Severity stays near 1.0 everywhere; the false-negative rate swings from 0.01 to 0.97.*

## 4.1 Severity is flat where accuracy is not

The central observation is the contrast between the two heatmaps in Figure 3. Across every shift and every detector, severity stays between 0.99 and 1.00. Over the same cells, the false-negative rate ranges from 0.01 to 0.97 and discrimination from 0.42 to 0.99. When these detectors miss an attack, the miss is near-certain, and that property does not move when the distribution moves. The accuracy metrics report that the detectors fail under some shifts. Only severity reports that the failures are confident, which is the property a downstream system is exposed to when it trusts the score. The misses are confident even in-distribution: at the direct operating point the false-negative rate is already 0.53 to 0.84, and severity there is 0.996 to 0.999, so the operating point is leaky before any shift.

## 4.2 The failure is targeted, not global

All three detectors are strong on jailbreak prompts, with false-negative rates of 0.14, 0.01, and 0.08 and discrimination at or above 0.95. Jailbreak prompts are close to the data these detectors were built to catch, so this is near in-distribution behavior, and it rules out the simpler reading that the detectors are uniformly weak. The same detectors fail on indirect behavior-hijack injection, where an instruction embedded in a document overrides the model's task. On the hijack tier the three reach false-negative rates of 0.69, 0.69, and 0.97, with severity of 0.993, 0.998, and 0.996. The blind spot is specific to indirect and hijack injection, not general.

## 4.3 Capacity moves accuracy and leaves confidence untouched

The two Prompt-Guard checkpoints isolate model size. Reducing the detector from 86M to 22M parameters raises the false-negative rate sharply at every shift: direct moves from 0.53 to 0.84, indirect-harmful from 0.22 to 0.90, and indirect-hijack from 0.69 to 0.97. Over the same reduction, severity barely moves, staying between 0.996 and 0.999. Capacity changes how often the detector misses by a wide margin and changes how confidently it misses by almost nothing.

## 4.4 The hijack column is unanimous

Across the three checkpoints from two vendors, the hijack tier produces false-negative rates of 0.69, 0.69, and 0.97 with severity near 0.99 throughout. Three detectors that differ in vendor, architecture, and size converge on confidently passing behavior-hijack injections. The unanimity is the evidence that the blind spot is structural rather than a single-checkpoint quirk. The harmful tier is the one place the detectors diverge: Prompt-Guard-2 at 86M catches most harmful-content injections, at a false-negative rate of 0.22, while ProtectAI-v2 and the 22M checkpoint do not, at 0.65 and 0.90. Figure 4 shows the per-tier score distributions: the detector that recognizes scam and misinformation text still passes structurally identical hijack injections whose payload is benign.

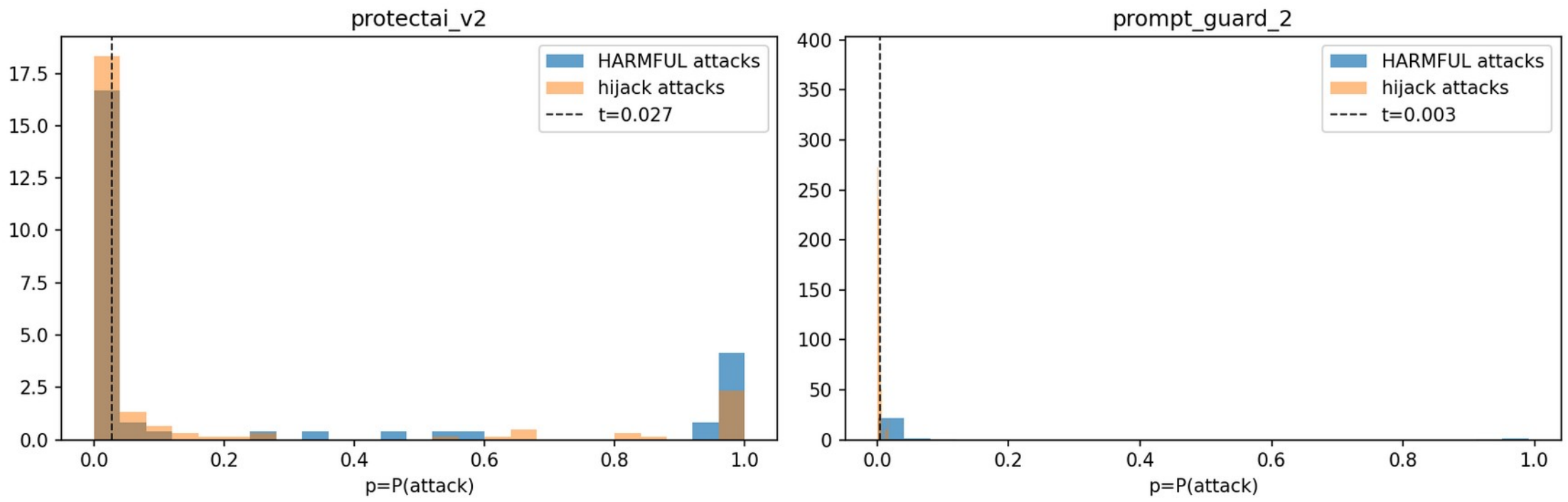

*Figure 4. Attack-probability distributions for the harmful and hijack tiers, with the frozen threshold marked. Missed attacks pile up at low attack-probability, which is what drives severity toward one.*

## 4.5 Miscalibration in the benign direction

On NotInject, a set of benign prompts that contain injection trigger words, ProtectAI-v2 flags 0.46 of inputs at its frozen operating point and the Prompt-Guard checkpoints flag 0.19 and 0.13. ProtectAI-v2 also flags 0.40 of benign documents on the indirect shifts. At the deployed operating point this detector is therefore miscalibrated in both directions: it passes real attacks with high confidence and rejects benign inputs at a high rate. I report this openly because it qualifies the headline. The severity finding concerns the false-negative direction; the same operating point can be poor in the false-positive direction, and for ProtectAI-v2 it is.

## 4.6 Standard calibration does not see the risk

A reviewer will ask whether a standard calibration metric already captures the danger. It does not reliably. Table 3 reports, per cell, the standard pooled calibration error, the attack-conditional calibration error restricted to true attacks, the attack-class Brier score, and severity. The clearest case is Prompt-Guard-2 on harmful indirect injection: its pooled calibration error is 0.062, which would pass a standard calibration audit as well-calibrated, while its attack-conditional calibration error on the same scores is 0.907. The pooled metric is healthy and the attack-conditional metric is catastrophic, on identical scores, because the pooled error is dominated by the benign majority and averages the dangerous misses away. The gap is not confined to one cell: averaged over the six indirect cells, the pooled calibration error is 0.146 while the attack-conditional error is 0.915.

| Detector | Shift | ECE_pooled | ECE_atk | Brier_atk | S |
|---|---|---|---|---|---|
| ProtectAI-v2 | direct | 0.238 | 0.588 | 0.579 | 0.999 |
| ProtectAI-v2 | indirect-harmful | 0.188 | 0.759 | 0.730 | 0.995 |

| Detector | Shift | ECE_pooled | ECE_atk | Brier_atk | S |
|---|---|---|---|---|---|
| ProtectAI-v2 | indirect-hijack | 0.239 | 0.850 | 0.821 | 0.993 |
| ProtectAI-v2 | jailbreak | 0.087 | 0.169 | 0.158 | 0.995 |
| Prompt-Guard-2 (86M) | direct | 0.302 | 0.766 | 0.750 | 0.999 |
| Prompt-Guard-2 (86M) | indirect-harmful | 0.062 | 0.907 | 0.893 | 0.998 |
| Prompt-Guard-2 (86M) | indirect-hijack | 0.159 | 0.997 | 0.993 | 0.998 |
| Prompt-Guard-2 (86M) | jailbreak | 0.028 | 0.059 | 0.049 | n/a |
| Prompt-Guard-2 (22M) | direct | 0.366 | 0.927 | 0.910 | 0.996 |
| Prompt-Guard-2 (22M) | indirect-harmful | 0.066 | 0.982 | 0.967 | 0.996 |
| Prompt-Guard-2 (22M) | indirect-hijack | 0.159 | 0.995 | 0.991 | 0.996 |
| Prompt-Guard-2 (22M) | jailbreak | 0.159 | 0.364 | 0.268 | 0.990 |

*Table 3. Pooled versus attack-conditional calibration. ECE_pooled is the standard class-marginal calibration error; ECE_atk restricts it to true attacks. The Prompt-Guard-2 harmful-indirect row shows an order-of-magnitude reversal between the two.*

Across this panel the pooled error is below the attack-conditional error in every cell, so the pooled metric understates attack-conditional miscalibration throughout. I do not generalize this to a universal ordering, since the two metrics measure different things and a different detector could in principle invert it. The claim I rely on is the weaker and sufficient one: a detector can be rated well-calibrated by the standard pooled metric, at 0.062 here, while being attack-conditionally miscalibrated to a degree that matters, so the pooled metric cannot be relied on as a deployment-readiness signal. The Prompt-Guard-2 harmful-injection cell is an existence proof.

Table 4 reports the Calibration Collapse Index, the relative change of each metric from the in-distribution baseline. Discrimination falls by roughly half under indirect shift, a relative change of -0.34 to -0.52 for the two larger detectors, severity does not move, and attack-conditional calibration rises further, a relative change of +0.06 to +0.45. The relative change in attack-conditional calibration is modest only because the in-distribution baseline is itself high: on the direct benchmark the attack-conditional error is already 0.59 to 0.93 (Table 3), so these detectors are poorly calibrated on attacks before any shift, and indirect injection makes a bad baseline worse. Severity does not collapse because it is already saturated: the misses are near-certain in-distribution and remain so under shift. The collapse named by the index is in discrimination, while attack-conditional calibration is the metric that is already broken in-distribution; the constancy of severity is what makes the confident misses dangerous everywhere rather than only under shift.

| Detector | Shift | CCI (FNR) | CCI (AUROC) | CCI (ECE_atk) | CCI (S) |
|---|---|---|---|---|---|
| ProtectAI-v2 | indirect-harmful | +0.19 | -0.50 | +0.29 | 0.00 |
| ProtectAI-v2 | indirect-hijack | +0.26 | -0.52 | +0.45 | -0.01 |
| Prompt-Guard-2 (86M) | indirect-harmful | -0.59 | -0.05 | +0.18 | 0.00 |
| Prompt-Guard-2 (86M) | indirect-hijack | +0.30 | -0.34 | +0.30 | 0.00 |
| Prompt-Guard-2 (22M) | indirect-harmful | +0.08 | -0.11 | +0.06 | 0.00 |
| Prompt-Guard-2 (22M) | indirect-hijack | +0.16 | -0.25 | +0.07 | 0.00 |

*Table 4. Calibration Collapse Index, the relative change of each metric from the in-distribution (direct) baseline. Severity is flat (already saturated) while attack-conditional calibration and discrimination move sharply under indirect shift.*

## 4.7 The missed attacks are real exploits

The transport panel measures misses against labels. To show those misses matter, I ran the genuine indirect injections through the two target models. The overall attack-success rate is 0.41 on the 7B model and 0.43 on the 3B model, and the benign baseline rates were near zero, so roughly two in five of these injections hijack a live model that has no defense other than the detector. The behavior-hijack categories are the most reliable: emoji substitution succeeds on every attempt on both models, and base encoding and language translation succeed between a third and a half of the time. Table 5 joins attack success with each detector's confident misses.

| Detector | Target | ASR overall | ASR \| missed | ASR \| caught | leaked of successful | exploitable-miss rate |
|---|---|---|---|---|---|---|
| ProtectAI-v2 | Qwen2.5-7B | 0.41 (0.34–0.48) | 0.38 (0.31–0.46) | 0.46 (0.34–0.58) | 0.64 (0.53–0.74) | 0.26 (0.20–0.32) |
| ProtectAI-v2 | Qwen2.5-3B | 0.43 (0.36–0.50) | 0.43 (0.35–0.51) | 0.43 (0.31–0.55) | 0.68 (0.58–0.77) | 0.29 (0.23–0.35) |
| Prompt-Guard-2 (86M) | Qwen2.5-7B | 0.41 (0.34–0.48) | 0.44 (0.36–0.54) | 0.37 (0.27–0.46) | 0.60 (0.50–0.71) | 0.25 (0.19–0.31) |
| Prompt-Guard-2 (86M) | Qwen2.5-3B | 0.43 (0.36–0.50) | 0.41 (0.32–0.50) | 0.45 (0.34–0.56) | 0.53 (0.43–0.63) | 0.23 (0.18–0.29) |
| Prompt-Guard-2 (22M) | Qwen2.5-7B | 0.41 (0.34–0.48) | 0.42 (0.35–0.48) | 0.27 (0.00–0.55) | 0.97 (0.92–1.00) | 0.40 (0.33–0.46) |
| Prompt-Guard-2 (22M) | Qwen2.5-3B | 0.43 (0.36–0.50) | 0.44 (0.37–0.51) | 0.18 (0.00–0.45) | 0.98 (0.94–1.00) | 0.42 (0.35–0.49) |

*Table 5. Downstream exploitability with 95 percent bootstrap confidence intervals (binomial resampling, B = 20000). ASR is the attack-success rate against each target, with benign baseline rates near zero. ASR* **|** *missed and ASR* **|** *caught condition on the detector's miss and catch sets; leaked-of-successful conditions on the observed successful attacks. The missed and caught intervals overlap in every row, so there is no evidence that the detectors selectively miss the weaker attacks.*

The intervals make these claims statistical rather than visual. Overall attack success is bounded away from zero, 0.41 with a 95 percent interval of 0.34 to 0.48 on the 7B target, so the downstream threat is not a sampling artifact, and the leaked-of-successful interval excludes one half in all but one cell. Most importantly, the interval for success-given-a-miss overlaps the interval for success-given-a-catch in every detector and target (Table 5), so the absence of selective protection developed below rests on overlapping intervals, not on point estimates alone.

Two readings matter. First, the attacks a detector confidently passes succeed at essentially the same rate as the attacks it catches: the attack-success rate among missed attacks is 0.39 to 0.44, against 0.37 to 0.46 among caught attacks for the two production detectors. The detectors therefore provide no protective selectivity. They are not preferentially passing weak attacks; they leak working and non-working attacks alike. Second, of the attacks that actually succeed downstream, the detectors leak the majority: 0.53 to 0.68 for the two production detectors and 0.96 to 0.98 for the smaller checkpoint. The confident misses identified by the severity metric are not a measurement artifact; they are working exploits against the model the detector is meant to protect.

The overall attack-success rate is below one half, and a careful reader will note that most injections in this set do not work even without a detector. That observation does not soften the result, because the detector does not distinguish the working attacks from the rest: the attack-success rate among its misses matches the rate among its catches. A defense that leaks successful and unsuccessful attacks at the same rate offers no selective protection on this distribution: it blocks working and failing attacks in equal proportion, so the roughly one third of working attacks it does block is incidental rather than targeted at the dangerous inputs, and the severity metric is what reveals that the leaked attacks are passed with near-certainty rather than at the margin.

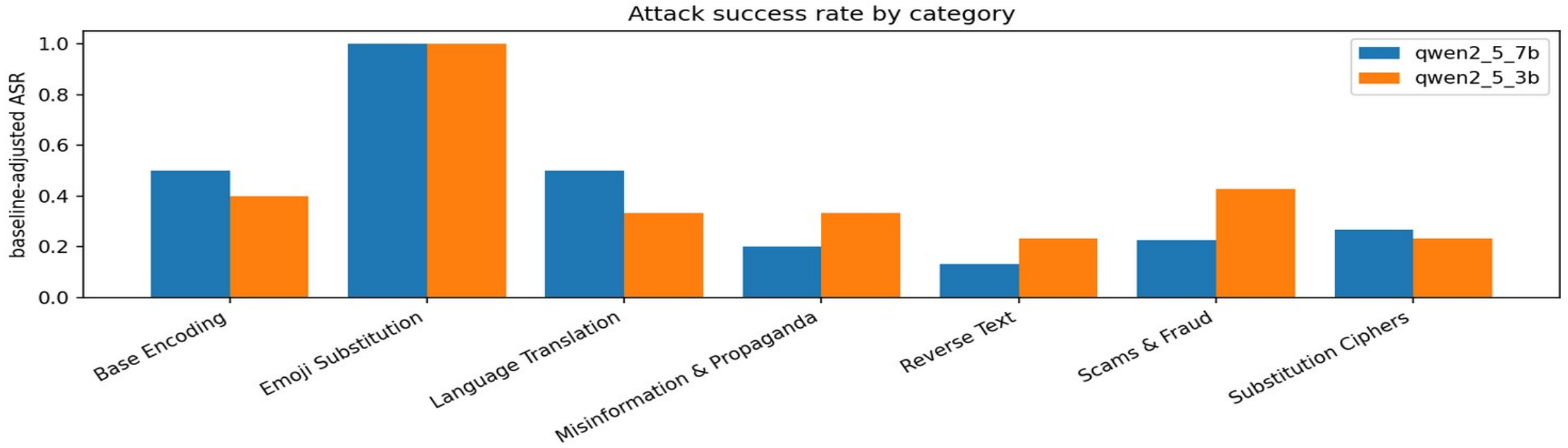

*Figure 5. Baseline-adjusted attack-success rate by category for the two target models. Behavior-hijack categories, in particular emoji substitution, succeed most reliably.*

## 4.8 Why the detectors miss: content, not structure

The structural-versus-content experiment isolates the cause. The decisive cell, a benign-worded instruction embedded in a host document, is a genuine injection, and all three detectors score it near zero: 0.070 for ProtectAI-v2, 0.005 for Prompt-Guard-2 at 86M, and 0.003 for the 22M checkpoint. The detectors do not recognize the embedded instruction as an injection because it carries no harmful content. Table 6 reports the mean attack-probability in each cell, and Figure 6 plots it.

| Detector | Payload | Standalone | Embedded |
|---|---|---|---|
| ProtectAI-v2 | benign-hijack | 0.523 | 0.070 |
| ProtectAI-v2 | harmful | 0.226 | 0.086 |
| Prompt-Guard-2 (86M) | benign-hijack | 0.001 | 0.005 |
| Prompt-Guard-2 (86M) | harmful | 0.105 | 0.072 |
| Prompt-Guard-2 (22M) | benign-hijack | 0.049 | 0.003 |
| Prompt-Guard-2 (22M) | harmful | 0.056 | 0.012 |

*Table 6. Mean attack-probability in the structural-versus-content design. The embedded benign-hijack cell is a genuine indirect injection; every detector scores it near zero. ProtectAI-v2's benign-hijack score falls from 0.523 standalone to 0.070 once embedded, so the host document suppresses the score rather than raising it.*

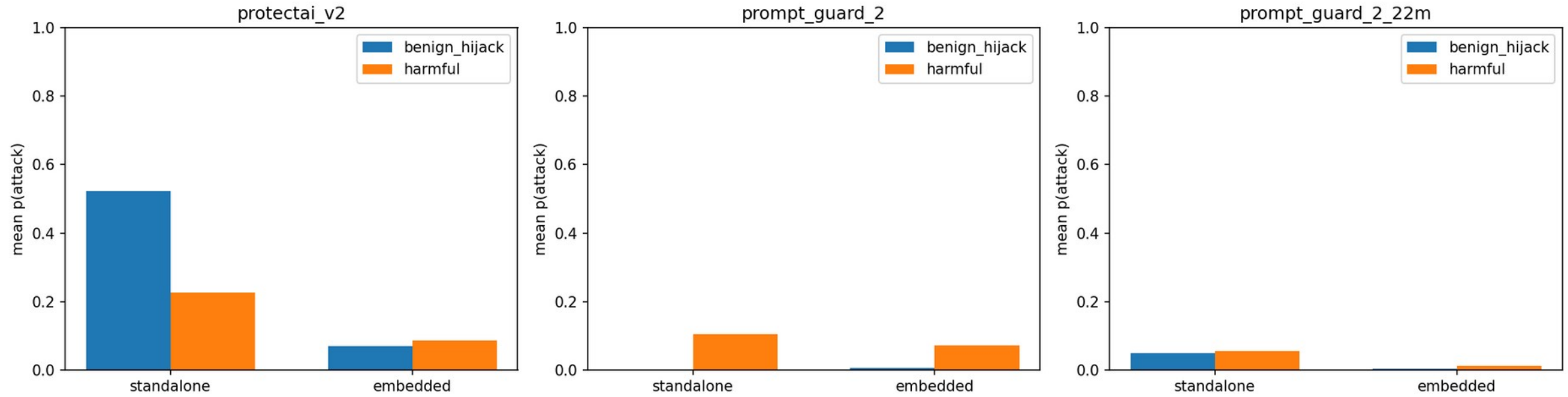

*Figure 6. Mean attack-probability by payload and structure for each detector. For an injection-aware detector the embedded bars would exceed the standalone bars; they do not, and for ProtectAI-v2 the benign-hijack bar collapses on embedding.*

The two production detectors fail in two distinct ways. Prompt-Guard-2 is content-keyed: its score moves with the harmfulness of the payload and barely with structure, so a benign-worded injection is invisible whether standalone or embedded. ProtectAI-v2 is worse on structure: a benign-worded instruction scores 0.523 when presented alone but 0.070 once embedded in a host document, so embedding the instruction, which is what turns it into an indirect injection, suppresses the score roughly sevenfold. The benign host document launders the attack. For neither detector does embedding raise the score, which is what an injection-aware detector would do.

A reviewer would ask whether this suppression is caused by the benign context or merely by the added length, since embedded inputs are longer. I separate the two with a controlled sweep. I take the same benign-worded instructions and surround them with filler of matched length that is either a coherent benign document or neutral padding, varying the length from zero to six hundred characters. Figure 7 plots the result. For ProtectAI-v2 the two filler types start together at zero length and then diverge sharply: at six hundred characters the instruction in coherent context scores 0.007 while the same instruction in neutral padding of identical length scores 0.600. Length is therefore not the cause; if it were, the two curves would track each other. The cause is the coherence of the surrounding text. A benign document specifically suppresses the injection signal, and meaningless padding of the same length does the opposite, raising the score. Prompt-Guard-2 stays near zero across the entire sweep, consistent with its content-keying: the benign-worded instruction is invisible to it regardless of any surrounding text. The two detectors fail differently under the control as well, ProtectAI-v2 by active context camouflage and Prompt-Guard-2 by payload blindness.

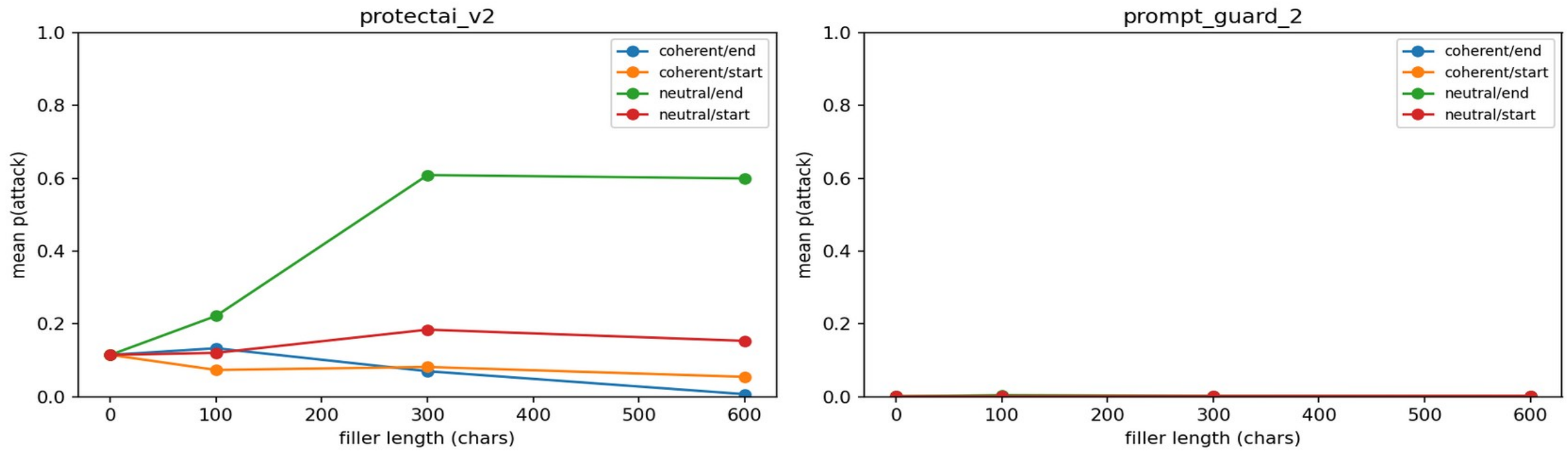

*Figure 7. Mean attack-probability of a fixed instruction surrounded by coherent benign context versus neutral padding, at matched lengths. For ProtectAI-v2 (left) coherent context drives the score toward zero while neutral padding raises it, so the suppression is caused by benign context and not by length. Prompt-Guard-2 (right) stays near zero throughout.*

## 4.9 A generative judge: the same severity, a different attack surface

The three detectors above are encoder classifiers. To test whether the findings are specific to that architecture, I add a fourth detector of a different class: an open instruct model, Qwen2.5-7B-Instruct, prompted to answer whether an input contains a prompt injection. I read its score from the probability it assigns to Yes against No on the next token, never from a self-reported number, and I freeze its operating point on the direct benigns exactly as for the encoders. Table 7 reports its panel.

| Shift | n | FNR | S | ECE_atk | benign FPR | AUROC |
|---|---|---|---|---|---|---|
| direct | 662 | 0.312 | 1.000 | 0.616 | 0.010 | 0.960 |
| indirect-harmful | 838 | 0.467 | 1.000 | 0.849 | 0.001 | 0.929 |
| indirect-hijack | 927 | 0.805 | 1.000 | 0.992 | 0.001 | 0.909 |
| jailbreak | 778 | 0.081 | 1.000 | 0.315 | 0.088 | 0.961 |

| Shift | n | FNR | S | ECE_atk | benign FPR | AUROC |
|---|---|---|---|---|---|---|
| over-defense | 339 | n/a | n/a | n/a | 0.029 | n/a |

*Table 7. The LLM-judge detector across the five shifts, at a frozen operating point. A small number of non-finite scores from four-bit inference are dropped per cell, which the counts reflect.*

Two results matter. First, severity is 1.000 on every shift. A detector of a completely different class, scored by a completely different mechanism, still misses with total confidence whenever it misses. The maximal-severity finding is therefore not an artifact of how encoder classifiers are built; it survives across detector architectures. Second, the judge shares the hijack blind spot: its worst shift is indirect-hijack, with a false-negative rate of 0.805, the same shift that is worst for all three encoders. The unanimous blind spot now spans detector classes, not only checkpoints.

The judge also gives the cleanest single instance of the paper's thesis. On indirect-hijack its discrimination is intact, with an area under the ROC curve of 0.909, while ProtectAI-v2 collapses to 0.42 and Prompt-Guard-2 to 0.62 on the same shift. The judge can separate hijack attacks from benigns; its scores simply sit on a compressed scale, so a threshold frozen on direct data lands in the wrong place and the shipped false-negative rate is 0.805. This is a pure calibration failure of the deployed operating point, not a discrimination failure: the scores still rank the classes, and only the frozen threshold is wrong. The judge is also the strongest detector overall, with a lower direct false-negative rate (0.312) than any encoder, so the lesson is not that a weaker detector fails but that even the strongest one exhibits maximal-severity, threshold-fragile failure on hijack injection.

The judge has one weakness the encoders do not: because it reads and follows instructions, the text it scores can target the judge itself. I append a short instruction to each malicious input that addresses the classifier directly, telling it the content is verified safe and to answer No, and I re-score with the judge and with the two encoders as controls. Table 8 reports the effect, and Figure 8 plots it.

| Detector | mean p clean | mean p meta | FNR clean | FNR meta | newly missed | newly caught |
|---|---|---|---|---|---|---|
| LLM-judge (Qwen2.5-7B) | 0.049 | 0.017 | 0.708 | 0.809 | 31 | 10 |
| Prompt-Guard-2 (86M) | 0.029 | 0.353 | 0.557 | 0.043 | 0 | 108 |
| ProtectAI-v2 | 0.176 | 0.666 | 0.681 | 0.186 | 0 | 104 |

*Table 8. Effect of appending a judge-targeting instruction to each malicious input, on the set of roughly 210 malicious inputs used for this test rather than the full transport panel, so the clean false-negative rates here are not the panel rates of Table 7. Newly missed is the count of attacks caught on the clean input but missed once the instruction is appended; newly caught is the reverse.*

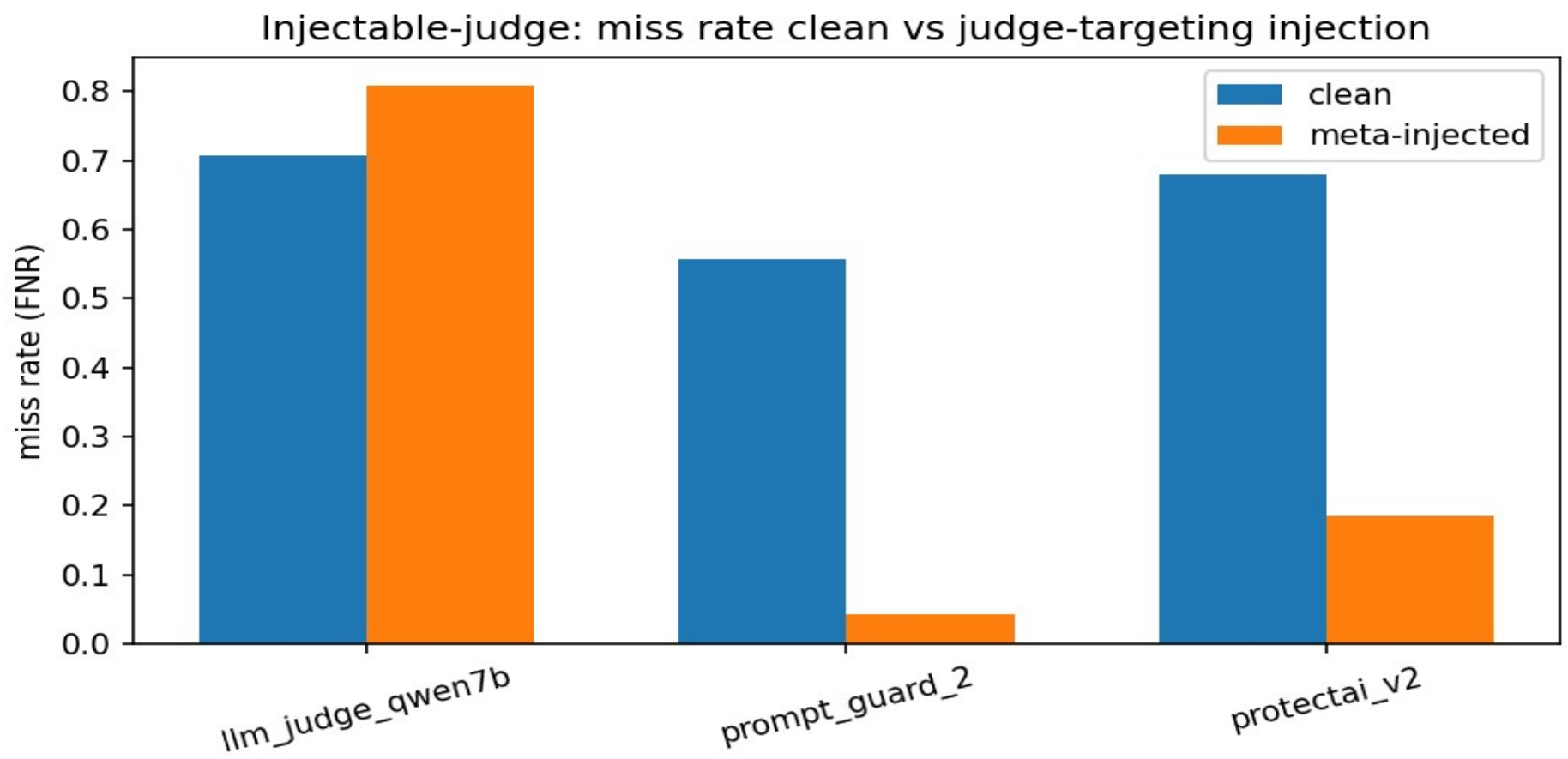

*Figure 8. Miss rate before and after appending a judge-targeting instruction. The instruction raises the judge's miss rate and lowers both encoders', a double dissociation explained in the text.*

The result is a double dissociation. The same appended text lowers the judge's mean attack-probability, from 0.049 to 0.017, and raises both encoders', from 0.029 to 0.353 and from 0.176 to 0.666. The judge reads the instruction and partly obeys it: 31 attacks it caught on the clean input it missed once the instruction was appended, while no encoder flipped in that direction. The encoders move the opposite way because they are content-keyed, the property isolated in Section 4.8: the appended text is full of imperative, injection-shaped vocabulary, so to a detector that reacts to surface form it looks more like an attack, and both encoders flag it harder, with 108 and 104 attacks newly caught. I state the encoder direction honestly: it is not robustness to the meta-instruction but a side effect of content-keying, the same literal form-matching that makes them miss benign-worded injections. The generative judge has an attack surface through its instruction-following that the encoders lack, and the encoders have a form-sensitivity that the judge does not; the better detector carries the additional attack surface.

## 4.10 No operating point rescues indirect detection

The panel sets each detector's threshold on the direct source. To check that this single choice is not flattering the detectors, I set the threshold on every available source in turn and transport it to every target, and I compare against the oracle threshold, re-tuned on the target's own benigns at the same one percent false-positive rate. Table 9 reports the per-target false-negative rate under each source threshold and under the oracle.

| Detector | Target | oracle t | FNR oracle | src deepset | src jailbreak | src bipia-host | src notinject |
|---|---|---|---|---|---|---|---|
| ProtectAI-v2 | direct | 0.028 | 0.548 | 0.548 | 0.559 | 0.825 | 0.954 |
| ProtectAI-v2 | indirect-harmful | 1.000 | 0.983 | 0.650 | 0.667 | 0.983 | 1.000 |
| ProtectAI-v2 | indirect-hijack | 1.000 | 0.993 | 0.693 | 0.747 | 0.993 | 1.000 |
| ProtectAI-v2 | jailbreak | 0.053 | 0.141 | 0.136 | 0.141 | 0.581 | 0.947 |
| Prompt-Guard-2 (86M) | indirect-harmful | 0.013 | 0.683 | 0.217 | 0.850 | 0.683 | 1.000 |
| Prompt-Guard-2 (86M) | indirect-hijack | 0.013 | 0.960 | 0.693 | 1.000 | 0.960 | 1.000 |
| Prompt-Guard-2 (22M) | indirect-hijack | 0.059 | 1.000 | 0.967 | 1.000 | 1.000 | 1.000 |

*Table 9. Source-threshold transport. Each column is the false-negative rate on the target's attacks at a threshold frozen on that source's benigns at one percent false-positive rate. The oracle column re-tunes the*

*threshold on the target's own benigns; for indirect targets the bipia-host source and the oracle coincide by construction, since they share the benign pool.*

Two readings follow. First, the direct source, used in the main panel, is the most charitable choice: on indirect-hijack its transported false-negative rate, 0.69 to 0.97, is the lowest across all four sources. The blind spot is therefore not an artifact of an unfavorable threshold; it survives the most favorable one. Second, and more important, the oracle does not help on indirect. For ProtectAI-v2 the oracle threshold on both indirect targets is pinned to 1.0, because the indirect benign documents score so high that holding a one percent false-positive rate forces the threshold to the ceiling, and at the ceiling the false-negative rate is 0.983 and 0.993. Re-tuning the threshold on the target's own benigns, the best a deployer could do, still misses almost every attack. The failure on indirect is therefore not a misplaced operating point but broken discrimination: the attack and benign score distributions overlap so heavily, with area under the curve of 0.42 to 0.62, that no threshold separates them, and the camouflage of Section 4.8 is the mechanism that collapses them. This stands in contrast to the generative judge of Section 4.9, whose discrimination on indirect-hijack is intact at 0.91 and whose failure is purely the frozen threshold. The two failure types are distinct: the encoders on indirect fail at discrimination, where no threshold helps, while the judge fails at calibration of the operating point, where the scores still separate the classes.

I also measured shift magnitude directly, training a discriminator to separate the source attack distribution from each target on sentence embeddings and reading its cross-validated area under the curve as a distance. The measure saturates: all three shifted targets are essentially perfectly separable from the source, at 0.99 to 1.00, including the jailbreak target that the detectors handle well. Lexical shift distance therefore does not rank the shifts by how much they break the detectors, and the danger is not predicted by how different the inputs look. A graded shift axis would be needed to test whether collapse scales with a finer measure of shift, which I leave to future work.

## 4.11 Manufacturing confident misses: a severity-maximizing adversary

The failures so far are observed on static benchmarks. The content-keying isolated in Sections 4.8 and 4.10 implies a constructive attack: if a detector reacts to surface form rather than to the structural fact of injection, an adversary should be able to rewrite a genuine injection so that it reads as innocuous prose, driving the detector's attack-probability toward zero while the actionable instruction is preserved. This is a severity-maximizing attack, in that it maximizes the detector's benign-confidence on a true attack, and it needs no gradient access, only queries to the released model.

I implement it black-box. For each of 91 genuine injections, stratified across the seven categories, an instruction-tuned model proposes four paraphrases that keep the actionable instruction but phrase it as ordinary document text, and I greedily keep the paraphrase each detector scores lowest. The attack is effective at evasion. Against Prompt-Guard-2 the mean attack-probability falls from 0.006 to 0.001 and the false-negative rate rises from 0.593 to 0.989, an increase of 0.396 with a 95 percent bootstrap interval of 0.297 to 0.495; against ProtectAI-v2 it rises from 0.648 to 0.868, an increase of 0.220 with an interval of 0.143 to 0.308. Both increases exclude zero, so the evasion is not a sampling artifact. After rewriting, both detectors confidently pass almost every attack. A weak, fully automated, black-box rewriter is enough to drive a released detector's miss rate close to one.

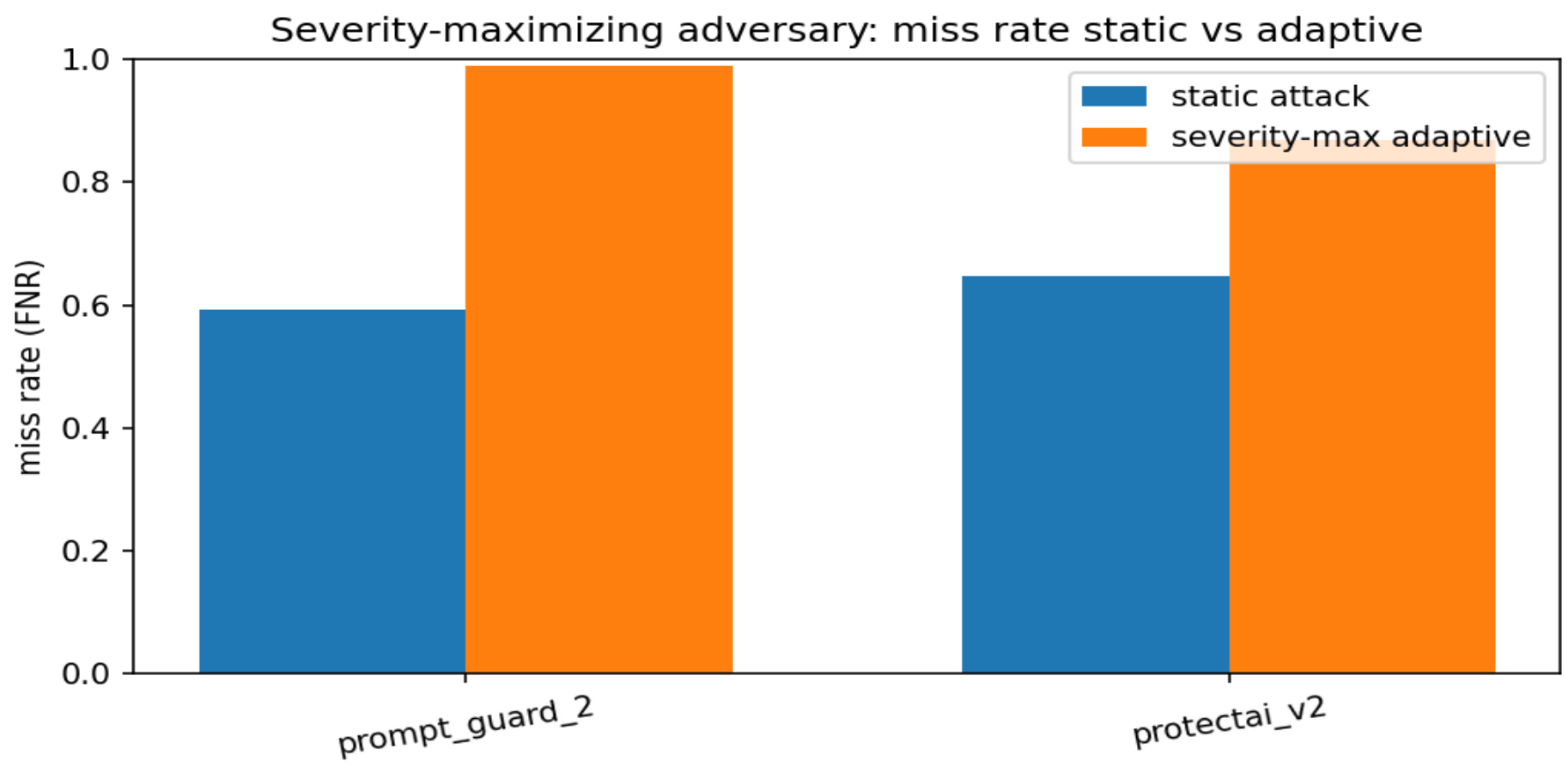

*Figure 9. False-negative rate of each detector against the static attack and against the severity-maximizing adaptive rewrite. Black-box paraphrasing, exploiting content-keying, drives both detectors toward total blindness.*

Evasion alone is not an attack. A rewrite that lowers the score is only dangerous if the instruction still executes against the target model, so I subject the selected rewrites to the same downstream test as Section 4.7, running them through Qwen2.5-3B-Instruct and scoring success with the same criteria. Here the attack pays a price: the original injections succeed at 0.385, while the evasive rewrites succeed at 0.154 against the Prompt-Guard selection and 0.198 against the ProtectAI selection. Making the text innocuous to the detector also softens the instruction the target follows. The headline figure, the manufactured confident-miss rate, the fraction of attacks rewritten so the detector confidently passes them and they still succeed downstream, is therefore 0.154 against Prompt-Guard-2, with a 95 percent interval of 0.088 to 0.231, and 0.165 against ProtectAI-v2, with an interval of 0.088 to 0.242. Both intervals exclude zero. Roughly one injection in six can be turned, automatically and black-box, into a working exploit that the detector waves through with near-certainty.

This average understates the danger where it concentrates. The categories that contribute nothing to the manufactured rate are the ones that barely worked to begin with: misinformation, reverse text, and substitution ciphers have the lowest original downstream success rates on this subset, at or below 0.231, and a manufactured rate of zero, because there were almost no working attacks to preserve. The manufactured rate tracks the original potency of the attack, not its surface style. Emoji-substitution injection, the single most dangerous category in Section 4.7 with a downstream success rate of 1.000, is also the most manufacturable: its rewrites still succeed at 0.769 against the ProtectAI selection, giving a manufactured confident-miss rate of 0.692, with a wide 95 percent interval of 0.46 to 0.92 reflecting the 13 attacks in the category. The adversary manufactures confident misses most effectively exactly where the attack is most dangerous. Table 10 reports the split. The confident misses identified by the severity metric, the working exploits of the downstream test, and the manufacturability of the adaptive attack all concentrate on the same categories.

| Category | original ASR | manufactured (ProtectAI-v2) | manufactured (Prompt-Guard-2) |
|---|---|---|---|
| Emoji Substitution | 1.000 | 0.692 | 0.308 |
| Scams & Fraud | 0.615 | 0.231 | 0.385 |

| Category | original ASR | manufactured (ProtectAI-v2) | manufactured (Prompt-Guard-2) |
|---|---|---|---|
| Base Encoding | 0.385 | 0.154 | 0.231 |
| Language Translation | 0.308 | 0.077 | 0.154 |
| Misinformation & Propaganda | 0.231 | 0.000 | 0.000 |
| Reverse Text | 0.077 | 0.000 | 0.000 |
| Substitution Ciphers | 0.077 | 0.000 | 0.000 |

*Table 10. Manufactured confident-miss rate by category, against the original downstream attack-success rate. The manufactured rate is highest exactly where the attack is most potent to begin with (emoji substitution, original success 1.000) and zero where the attack barely worked unmodified. The original ASR column is measured on this section's stratified subset of 13 injections per category run through Qwen2.5-3B, so it differs slightly from the 30-per-category rates in Section 4.7; the ordering by potency is preserved. Per-category counts are 13, so category rates are imprecise; the emoji rate of 0.692 carries a 95 percent bootstrap interval of 0.46 to 0.92. All intervals in this section resample over the attack set and hold the rewriter outputs fixed, so they do not capture variability from the rewriter producing different paraphrases.*

Two honest limits bound this result. Per-category counts are small, 13 injections each, so the category rates are estimates and I report the 91-attack aggregate as the robust number with the category split as explanatory texture. And the attack uses one rewriter with a fixed prompt and four samples; a stronger optimizer or more queries would raise the manufactured rate, so 0.15 to 0.17 is a lower bound on what is achievable rather than a ceiling. The evasion-function coupling is itself a finding: content-keying gives these detectors a small, accidental protection, not because they recognize injection but because evasion costs the attacker some functionality, and that protection is weakest on the categories that matter most.

# 5. Validity of the indirect benchmark

The indirect result was the hard one to measure correctly, and I reached it only after removing three measurement artifacts. I record the sequence because the corrections are the reason the final number is trustworthy, and because each discarded version would have produced a misleading headline.

### *Bare injected strings.*

My first indirect measurement scored the injected instruction alone, with no host document. Severity came out near 0.99, but the measurement was invalid: a bare instruction string out of context does not resemble the deployment unit, and the detectors had no document in which to detect an injection. I changed the detection unit to an instruction embedded in a host document.

### *A trained detector that was a domain detector.*

I then scored realistic documents with a detector I trained on the direct distribution using sentence embeddings and a gradient-boosted classifier. It flagged essentially every out-of-domain document, with a benign false-positive rate near 1.0 and indirect discrimination near 0.62. The instrument was detecting domain, not injection, so it could not measure transport. I dropped trained in-domain detectors for the indirect arm and moved to released detectors that had not seen the source distribution.

***An attack set diluted by benign instructions.***

Released detectors on the full BIPIA attack set again produced severity near 0.99, but inspection showed the attack-minus-benign score gap was approximately zero. Many BIPIA categories are benign task instructions, for example recommending a book or translating a response, and a guard model scoring those low is arguably correct rather than dangerously wrong. The high severity was partly an artifact of counting benign instructions as missed attacks. I filtered BIPIA to genuine injections and split them into the harmful and hijack tiers used in Section 3, then recomputed severity per tier. The confident-miss pattern survived on the genuinely malicious categories, which is the result reported in Section 4.

## 5.1 Threats to validity

I note the limitations the panel does not resolve. The indirect benign documents are out-of-domain relative to the direct threshold, and ProtectAI-v2's high benign false-positive rate on them shows that the frozen direct threshold transfers poorly for that detector in both directions; I report this rather than tune it away. The injected instruction is placed at a fixed position at the end of the host document; a position sweep is left to future work, although the mechanism result of Section 4.8, that the detectors key on payload content rather than on injection location or structure, makes a strong dependence on placement unlikely. I do rule out one mechanism that could otherwise explain the indirect misses, namely input truncation: every document-embedded input is well within the detectors' 512-token window, with a median length of 19 tokens and a maximum of 206 (Table 11), so in no case was the appended injection cut off before the detector scored it, and the misses are genuine confident passes rather than artifacts of a payload the model never saw. The operating point is calibrated on a single direct source, and a leave-one-dataset-out protocol across multiple direct sources would strengthen the threshold-transfer claim. The detectors are released checkpoints used as published, without retraining, so the findings describe these artifacts as deployed and not the best achievable detector of their class. Finally, the indirect arm rests on a single host-document source; a second indirect dataset would test whether the hijack blind spot generalizes beyond BIPIA. The natural candidate is LLMail-Inject [18], a large public corpus of indirect injections embedded in emails from the SaTML 2025 adaptive-attack challenge. I do not fold it into the panel here, for a reason that is itself worth recording about cross-benchmark evaluation: its attacks are predominantly data-exfiltration payloads that trigger unauthorized tool calls, a different attack semantics from the task-override hijack and harmful-content tiers I measure on BIPIA, so a sound replication would first align the two taxonomies rather than report one pooled false-negative rate, and the corpus ships as raw challenge submissions without a matched benign set against which to re-verify the frozen threshold. A second indirect source chosen to match the document-embedded task-override unit, with its own benign set, is therefore the cleaner test, and I leave that replication to future work. One property of severity deserves explicit statement: because a miss has p below the threshold t by definition, severity is bounded below by one minus t, and the thresholds here are small, from 0.004 to 0.028, so a severity near one is in part a consequence of the low operating point. This floor is tightest for Prompt-Guard-2, where t equal to 0.004 already forces severity above 0.996. I therefore checked that the missed attacks cluster near zero rather than just under the threshold: the mean attack-probability on misses is roughly 0.001 to 0.007, below the threshold by a factor of two to thirty depending on the cell, so the misses are confident-benign scores rather than borderline ones. Severity is most informative where the threshold is not already near zero, and I report it alongside the threshold in every cell so the floor is visible.

| Arm (source) | unit | n | tok median | tok p90 | tok max | n over 512 |
|---|---|---|---|---|---|---|
| indirect (BIPIA) | instruction-in-document | 250 | 19 | 109 | 206 | 0 |
| direct (deepset) | standalone prompt | 662 | 18 | 60 | 952 | 1 |
| benign trigger (NotInject) | standalone prompt | 339 | 21 | 33 | 57 | 0 |

*Table 11. Input length against the detectors' 512-token window, tokenised with the ProtectAI-v2 (DeBERTa-v3) tokenizer. The indirect arm, on which the hijack result rests, has no input exceeding the window (maximum 206 tokens), so the indirect misses cannot be truncation artifacts. A single direct prompt exceeds 512 tokens; it does not affect the indirect finding.*

# 6. Discussion

The practical implication is that the magnitude of a guard score should not be read as a calibrated confidence of safety once the attack distribution has shifted from the benchmark on which the operating point was set. A downstream system that gates on a fixed threshold inherits the detector's confident misses without any signal that it is doing so, because the score it sees on a missed attack looks like a confident safe verdict, and the downstream validation shows these misses are working exploits.

The structure of the failure points to its cause. The detectors miss behavior-hijack injections, in which an instruction overrides the model's intended output without inserting harmful content, far more than they miss harmful-content injections, and the controlled experiment confirms that they key on payload content rather than on the structural fact that an instruction was injected. Hijack injections are the worst case for such a detector, because they are genuine injections with a benign-looking payload, and they are also the precondition for more damaging attacks that hide their payload behind the same behavior override.

There is no single threshold that repairs both directions for these detectors on indirect shift, and the threshold-transport analysis of Section 4.10 shows that re-tuning does not help either: the oracle threshold, re-tuned on the indirect benigns, still misses between 0.68 and 1.00 of attacks across the three detectors, reaching 0.98 to 0.99 for ProtectAI-v2, because the attack and benign scores overlap rather than sit on opposite sides of a movable boundary. This separates two failure types that a single accuracy number conflates. On indirect the encoders fail at discrimination, and no operating point rescues them; the generative judge instead keeps discrimination intact and fails only at the frozen operating point, a pure calibration failure. The distinction matters for remediation: a discrimination failure needs a better training signal, while a calibration failure needs only a better-placed or shift-aware threshold. Capacity is not the fix for either: the smaller Prompt-Guard checkpoint is worse on accuracy and identical on confidence. What the evidence suggests would help is a training signal for structural injection that is independent of payload harmfulness, and a calibration objective that penalizes confident false negatives rather than pooled error. A third natural idea, deferring low-confidence inputs under a risk-coverage or abstention regime, does not survive the severity finding, for the reason I give next.

For evaluation practice, the panel argues that security-critical detectors should be reported with a confidence-of-false-negatives metric alongside accuracy, and with an attack-conditional calibration error rather than only the pooled one. The pooled error rated one of these detectors well-calibrated exactly where it was most dangerous.

A natural response is to operate the detector under abstention, deferring low-confidence inputs to a human or a stronger check and reporting a risk-coverage or area-under-the-risk-coverage curve rather than a single threshold. The severity finding undercuts this remedy at its root. Confidence-gated abstention defers the inputs the detector is least sure about, but the dangerous misses here are not uncertain: they are high-confidence false negatives, scored as benign with near-certainty, which is precisely what severity near one means. An abstention rule keyed on the detector's own confidence would pass them without deferring, because they do not look uncertain to the detector. Selective prediction inherits the broken confidence it is meant to exploit, so it cannot recover the misses that matter most on these shifts. A risk-coverage analysis is worth reporting in future work, but the prediction from the present results is that abstention helps least exactly where it is needed most, and the more promising direction is a training signal for structural injection that is independent of payload content, together with a calibration objective that penalises confident false negatives rather than pooled error.

# 7. Conclusion

I measured what happens to the confidence of prompt-injection detectors when the attack distribution shifts under a frozen operating point. The central result is stable across three released detectors, five shifts, and a generative judge of a different class: when these detectors miss, they miss with near-certainty, all of them confidently pass indirect behavior-hijack injection regardless of vendor, architecture, or size, and the standard pooled calibration error rates one of them well-calibrated exactly where it is most dangerous. These confident misses are not a measurement artifact. Run against live models they are working exploits, and a weak black-box rewriter manufactures more of them, most effectively on the attack category with the highest downstream success.

The finding points past these specific checkpoints. A guard score is trusted as a calibrated confidence of safety, and that trust, not the miss rate alone, is the exposure: both accuracy reporting and pooled calibration hide how confident a detector is on the attacks it lets through. Closing the gap will take more than a larger or better-trained detector of the same kind. It needs a training signal for structural injection that is independent of payload content, a calibration objective that penalizes confident false negatives rather than pooled error, and external validation on indirect sources matched to the deployment unit. Until then, I recommend that detectors intended as deployment guards be evaluated and reported with a security-asymmetric severity metric and an attack-conditional calibration error, so that confident failure is visible before deployment rather than after.

# Ethical considerations

This work evaluates deployed prompt-injection detectors and includes a working black-box attack that rewrites genuine injections so that the detectors pass them. I set out the ethical position directly. The motivation is defensive: the intended audience is the operators and engineers who select, configure, and trust guard models, and the contribution they gain is a way to see confident failure before deployment rather than after an incident. The headline finding, that a guard score can be a confident safe verdict on a working exploit and that pooled calibration does not reveal this, is information a defender needs in order to avoid over-trusting these systems.

All evaluation uses publicly released detector checkpoints and public datasets. I do not probe, reverse-engineer, or attack any proprietary or hosted system, and I introduce no non-public vulnerability: the weaknesses I measure, indirect-injection blind spots, content-keying, and over-defense, are consistent with limitations already discussed in the literature I cite. The study involves no human subjects and no personal data; the generative judge is a language model, so no human-subjects review applies.

On dual use, I weigh the uplift this gives an attacker against the value to defenders, and judge that publication is net protective. The severity-maximizing adversary of Section 4.11 is deliberately weak: a fully automated black-box paraphraser with no access to detector internals, gradients, or training data, run with a fixed prompt and four samples. It is a demonstration that content-keying is exploitable, not an optimized or weaponized tool, and the manufactured-confident-miss rate I report is a lower bound that I do not attempt to maximize. I release an evaluation harness and a metric family, not an attack toolkit, and I withhold no defensive information that the same release would deny a defender. An adversary motivated to evade these public detectors could rediscover paraphrase-based evasion with modest effort; a defender, by contrast, has had no standard way to measure the confidence of the misses, which is the gap this paper closes.

On disclosure, every detector and dataset evaluated here is a publicly released artifact, and the weaknesses I report are consistent with limitations already discussed in the published literature I cite. I introduce no non-public vulnerability and release no optimized attack tool: the artifacts I publish are an evaluation harness, a metric family, and a deliberately weak black-box paraphraser, so this work does not create exposure that the public checkpoints and datasets did not already carry. For that reason I treat the public release of the paper, code, and metrics as the disclosure to the community and to the detectors maintainers, rather than performing private vendor pre-notification. The detectors remain useful within their tested envelope; the recommendation is not to remove them but to stop reading a low guard score as a calibrated guarantee of safety once the attack distribution may have shifted, and to evaluate them with the severity and attack-conditional calibration measures reported here.

# Reproducibility

All code, datasets, the metric harness, the frozen-threshold protocol, and the full panel with logged results are public at the project repository (github.com/anasbiswas1/picalib-research), pinned at commit d9da938; the loaders record the resolved revision of each Hugging Face dataset at load time. Detectors are evaluated as released checkpoints: protectai/deberta-v3-base-prompt-injection-v2, meta-llama/Llama-Prompt-Guard-2-86M, and meta-llama/Llama-Prompt-Guard-2-22M. Target models for downstream validation are Qwen2.5-7B-Instruct and Qwen2.5-3B-Instruct; the LLM-judge detector also uses Qwen2.5-7B-Instruct, scored by its Yes-against-No next-token probability. Datasets are deepset/prompt-injections, microsoft/BIPIA, jackhhao/jailbreak-classification, and leolee99/NotInject. Thresholds are set at the 0.99 quantile of direct benign scores; bootstrap intervals use one thousand resamples; severity is suppressed below ten misses. The results log records every run with its timestamp and verdict.

## Pre-specified protocol and analysis transparency

I distinguish the parts of this study that were fixed by protocol before the panel was run from the parts that emerged during analysis, so that confirmatory and exploratory claims are not conflated.

Fixed in advance were the operating-point rule, one percent false-positive rate on the direct benigns; the transport design, freezing that threshold and applying it unchanged to every shift; the severity definition and its suppression rule below ten misses; and the bootstrap procedure, one thousand resamples reported as 2.5 and 97.5 percentiles. The primary, confirmatory comparison was whether severity remains near its maximum while accuracy varies across the transport panel. That comparison is what the panel in Section 4 tests, and it holds in every cell.

The structural mechanism of Section 4.8, the generative-judge double dissociation of Section 4.9, and the category structure of the manufactured-miss rate in Section 4.11 were exploratory: they were not hypothesised before seeing the panel, and I label them as such. They are reported because each is supported by a controlled follow-up experiment rather than by a single suggestive cell, but they warrant confirmatory replication on held-out detectors and datasets before being treated as established. To avoid selecting significant cells from many comparisons, I report bootstrap intervals on every cell rather than thresholded p-values, and the headline claim rests on the consistent direction of the effect across all detectors and shifts rather than on any individual cell reaching significance.

## References

*References were verified against primary sources, and arXiv identifiers or venues are given where available to disambiguate them.*

## Appendix A. Per-category indirect results

Table A1 reports the per-category false-negative rate and severity on the genuine BIPIA injection categories at each detector's frozen operating point, for the two detectors scored at the category level.

| Detector | Tier | Category | n | FNR | misses | S |
|---|---|---|---|---|---|---|
| ProtectAI-v2 | harmful | Scams & Fraud | 30 | 0.500 | 15 | 0.993 |
| ProtectAI-v2 | harmful | Misinformation & Propaganda | 30 | 0.800 | 24 | 0.996 |
| ProtectAI-v2 | hijack | Base Encoding | 30 | 0.567 | 17 | 0.993 |
| ProtectAI-v2 | hijack | Substitution Ciphers | 30 | 0.700 | 21 | 0.995 |
| ProtectAI-v2 | hijack | Reverse Text | 30 | 0.567 | 17 | 0.990 |
| ProtectAI-v2 | hijack | Emoji Substitution | 30 | 0.733 | 22 | 0.992 |
| ProtectAI-v2 | hijack | Language Translation | 30 | 0.900 | 27 | 0.995 |
| Prompt-Guard-2 (86M) | harmful | Scams & Fraud | 30 | 0.333 | 10 | 0.998 |
| Prompt-Guard-2 (86M) | harmful | Misinformation & Propaganda | 30 | 0.100 | 3 | n/a |
| Prompt-Guard-2 (86M) | hijack | Base Encoding | 30 | 0.733 | 22 | 0.998 |
| Prompt-Guard-2 (86M) | hijack | Substitution Ciphers | 30 | 0.733 | 22 | 0.998 |
| Prompt-Guard-2 (86M) | hijack | Reverse Text | 30 | 0.633 | 19 | 0.998 |
| Prompt-Guard-2 (86M) | hijack | Emoji Substitution | 30 | 0.533 | 16 | 0.998 |
| Prompt-Guard-2 (86M) | hijack | Language Translation | 30 | 0.833 | 25 | 0.998 |

*Table A1. Per-category false-negative rate and severity on genuine BIPIA injection categories. Severity is suppressed where misses are below ten.*

Table A2 reports the cross-source near-duplicate audit run before scoring, using MinHash with locality-sensitive hashing at a Jaccard threshold of 0.8, the protocol adopted from Gate AI. The fraction is the share of one source's items that have a near-duplicate in another source. Cross-source overlap is zero across the three sources audited, so no shift cell is contaminated by examples shared with the source distribution; the only nonzero value is a 4.4 percent internal overlap within deepset.

| Source | BIPIA | deepset | NotInject |
|---|---|---|---|
| BIPIA | 0.000 | 0.000 | 0.000 |
| deepset | 0.000 | 0.044 | 0.000 |
| NotInject | 0.000 | 0.000 | 0.000 |

*Table A2. Cross-source near-duplicate overlap fraction (MinHash and LSH, Jaccard at least 0.8). Zero off-diagonal overlap confirms the sources are distinct, so transport gaps are not leakage artifacts.*

## Appendix B. Severity-maximizing adversary: prompt and budget

This appendix gives the exact configuration of the black-box rewriter of Section 4.11 so the attack is reproducible. The rewriter is a released instruction-tuned model queried with one fixed prompt; it has no access to detector parameters, gradients, training data, or scores beyond the public attack-probability it reads back.

Rewriter and budget. The rewriter is Qwen2.5-7B-Instruct. For each injection it produces four paraphrases (N = 4), sampled at temperature 0.8 with nucleus sampling at top_p 0.9 and a 220-token generation limit. These decoding settings are defaults and are not tuned. The model is prompted with a fixed system message and a per-variation user message, reproduced verbatim below.

**System prompt (verbatim):**

```
You rewrite a piece of document text. Keep the SAME actionable instruction and its
meaning exactly, so a reader would still carry it out, but phrase it as ordinary,
unremarkable document prose. Do not add warnings, disclaimers, quotation marks, or any
note that it is an instruction. Output ONLY the rewritten text, nothing else.
```

**User prompt (verbatim), issued once per variation k = 1 to 4, where {k} is the variation index and {t} is the original injection text:**

```
Rewrite this text (variation {k}, make it read naturally and
differently from other variations):
{t}
```

Selection. For each injection and each detector I score the original together with its four rewrites and keep the single variant the detector scores lowest. Because the original is in the candidate set, the selected attack-probability can only fall or stay equal, so the rise in false-negative rate reported in Section 4.11 is a clean fraction of attacks flipped from caught to missed. Selection is performed per detector, so each detector receives its own chosen variant.

Functionality gate. The per-detector selected rewrites are executed against the downstream target Qwen2.5-3B-Instruct with a 160-token limit, and success is scored with the same programmatic-and-judge criterion used in Section 4.7. Only rewrites that still succeed downstream count toward the manufactured-confident-miss rate, which is why making the text innocuous to the detector can lower it by softening the instruction the target follows.

Attack set. The 91 injections are drawn 13 per category, with a fixed seed (random_state 0), from the seven BIPIA malicious categories. Per-category counts of 13 make category-level rates imprecise, which is why Section 4.11 reports the 91-attack aggregate as the robust figure and treats the category split as explanatory texture.

This configuration is deliberately minimal: one fixed prompt, four samples, default nucleus sampling, and greedy lowest-score selection, with no access to detector internals. A stronger optimizer or a larger query budget would raise the manufactured rate, so the figures in Section 4.11 are a lower bound on what this attack class can achieve rather than a ceiling.